\documentstyle[floats,prl,aps]{revtex}
\begin{document}
\renewcommand{\thefootnote}{\fnsymbol{footnote}}
\draft
\title{\large\bf 
Algebraic Bethe ansatz for
integrable Kondo impurities in the one-dimensional supersymmetric t-J model} 

\author { Huan-Qiang Zhou and Mark D. Gould } 

\address{      Department of Mathematics,University of Queensland,
		     Brisbane, Qld 4072, Australia}

\maketitle

\vspace{10pt}

\begin{abstract}
An integrable Kondo problem in the one-dimensional supersymmetric t-J
model is studied by means of the boundary supersymmetric quantum inverse
scattering method.
The boundary $K$ matrices depending on the local moments of the impurities 
are presented  as a nontrivial realization of the graded reflection equation
algebras in a two-dimensional impurity Hilbert space.
Further,the model is solved by using the algebraic Bethe ansatz method
and the Bethe ansatz equations are obtained. 
\end{abstract}




\def\a{\alpha}
\def\b{\beta}
\def\d{\delta}
\def\e{\epsilon}
\def\g{\gamma}
\def\k{\kappa}
\def\l{\lambda}
\def\o{\omega}
\def\t{\theta}
\def\s{\sigma}
\def\D{\Delta}
\def\L{\Lambda}


\def\beq{\begin{equation}}
\def\eeq{\end{equation}}
\def\bea{\begin{eqnarray}}
\def\eea{\end{eqnarray}}
\def\ba{\begin{array}}
\def\ea{\end{array}}
\def\no{\nonumber}
\def\le{\langle}
\def\re{\rangle}
\def\lt{\left}
\def\rt{\right}

\newcommand{\reff}[1]{eq.~(\ref{#1})}

\vskip.3in

Recently,much interest has been devoted to the investigation
of the theory of impurities coupled to a Luttinger liquid. 
Such a problem was first considered by Lee and Toner \cite {LT92}. By using the
perturbative renormalization group theory they found the crossover
of the Kondo temperature from power law dependence on the Kondo coupling
constant to an exponential one.  And then,a poor man's scaling was
carried out by Furusaki and Nagaosa \cite {FN94} ,who found
a stable strong coupling fixed point  for both antiferromagnetic
and ferromagnetic cases. On the other hand,boundary conformal field
theory predicts two types of critical behaviours ,i.e.,either a local 
Fermi liquid with standard low-temperature thermodynamics or a non-Fermi
liquid observed by Furusaki and Nagaosa \cite {FN94}. However,in order to get a full
picture about the critical behaviour
of Kondo impurities coupled to a Luttinger liquid,some simple
integrable models
which allow exact solutions are desirable.

Several integrable magnetic and nonmagnetic impurity problems
describing a few impurities embedded
in some correlated electron systems have so far appeared in the
literature. Among them are several versions of the supersymmetric $t$-$J$
model with impurities \cite {BAR94,BEF97,LF98}. Such an idea to incorporate an impurity into a
closed chain may date back to Andrei and Johanesson \cite {AJ84}(see
also \cite {ZJ89}). However,the model
thus constructed suffers the lack of backward scattering and results in
a very complicated Hamiltonian. Therefore, as observed by Kane and
Fisher \cite {KF92},it seems to be advantageous to adopt open boundary
conditions with
the impurities at open ends to study the Kondo impurities coupled to
integrable strongly correlated electron systems \cite {PW97}. 

In this letter, a formalism is presented to systematically construct
integrable Kondo problems in  strongly correlated electron systems. Our
new input is to search for integrable boundary $K$ matrices
depending on the local moments of impurities,which arise as a nontrivial
realization of
the graded reflection equation algebras. The formalism is applied to
the 1D supersymmetric t-J model,and the corresponding boundary K matrices are
constructed. It should be emphasized that our new non-c-number boundary 
K matrices are highly nontrivial, 
in the sense that they can not be factorized into the product of a
c-number boundary K matrix and the corresponding local monodromy
matrices. The model is solved by means of the 
algebraic Bethe ansatz method and
the Bethe ansatz equations are derived.

Let $c_{j,\s}^\dagger$ and $c_{j,\s}$ denote  creation and
annihilation operators of the conduction electrons with spin $\s$ at
site $j$, which satisfy the anti-commutation relations given by
$\{c_{i,\s}^\dagger, c_{j,\s'}\}=\d_{ij}\d_{\s\s'}$, where 
$i,j=1,2,\cdots,L$ and $\s,\s'=\uparrow,\;\downarrow$. We consider the
following Hamiltonian describing two magnetic impurities coupled to
the supersymmetric t-J open chain,
\bea
H&=&-\sum _{j=1, \s}^{L-1}{\cal P} (c^\dagger_{j\s}c_{j+1\s}+H. c.){\cal
P}+
  2 \sum_{j=1}^{L-1}({\bf S}_j \cdot {\bf S}_{j+1}-
  \frac {1}{4}n_jn_{j+1})+\no\\
& &J_a{\bf S}_1 \cdot {\bf S}_a+V_an_1+J_b{\bf S}_L \cdot {\bf S}_b
+V_b n_L,\label {ham}
\eea
where the projector ${\cal P}=\prod ^L_{j=1} (1-n_{j\uparrow}n_{j\downarrow})$
ensures that double occupancies of sites are forbidden; $J_\a,V_\a (\a=a,b)$ 
are the Kondo coupling constants and the
impurity scalar potentials respectively; ${\bf S}_j=\frac {1}{2}\sum
_{\s,\s'}c^\dagger_{j\s}{\bf \s}_{\s\s'}c_{i\s'}$ is the spin operator
of the conduction electrons; ${\bf S}_{\a} (\a = a,b)$ are the local
moments with spin-$\frac {1}{2}$ located at the left and right ends of
the system respectively;
 $n_{j\s}$ is the number density operator
$n_{j\s}=c_{j\s}^{\dagger}c_{j\s}$,
$n_j=n_{j\uparrow}+n_{j\downarrow}$.

As was shown in Refs. \cite {SAK91,SCH87,BBO91}, the supersymmetry algebra underlying the bulk
Hamiltonian of this model is
$gl(2|1)$,and the integrability of the model on a closed chain has been
formulated by Essler and Korepin \cite {EK92} and Foerster and Karowski
\cite {FK92},independently. It is quite interesting to note that although
the introduction  of the impurities spoils the supersymmetry,there is
still a remaining  $su (2)$ symmetry in the Hamiltonian (\ref {ham}).
Below we will establish the quantum integrability of the Hamiltonian
(\ref{ham}) for a special choice of the model parameters $J_\a$ and
$V_\a$,
\beq
J_\a = -\frac {8}{(2c_\a -1)(2c_\a+3)},
V_\a = -\frac {4c_\a^2-3}{(2c_\a -1)(2c_\a+3)}.
\eeq
This is achieved by showing that it can be derived from
the (graded) boundary quantum inverse scattering method \cite
{Zhou97,BRA98}. Our result is consistent with the applicability of the
coordinate Bethe ansatz method \cite {PW97}.

Let us recall that the Hamiltonian of
the 1D  supersymmetric t-J model with the periodic boundary conditions
commutes with the transfer matrix, which is the supertrace of the
monodromy matrix $T(u)$,
\beq
T(u) = R_{0L}(u)\cdots R_{01}(u). \label{matrix-t}
\eeq
Here the quantum R-matrix 
$ R_{0j}(u)$ takes the form, 
\beq
R=u+P, \label {r}
\eeq
where $u$ is the spectral parameter,and $P$ denotes the graded
permutation operator,
and the subscript $0$ denotes the 3-D auxiliary superspace $V=C^{2,1}$ with
the grading $P[i]=1,(i=1,2)$ and $P[3]=0$.
It should be noted that the supertrace
is carried out for the auxiliary superspace $V$.
The elements of the supermatrix $T(u)$ are the generators
of an associative superalgebra ${\cal A}$ defined by the relations
\beq
R_{12}(u_1-u_2) \stackrel {1}{T}(u_1) \stackrel {2}{T}(u_2) =
   \stackrel {2}{T}(u_2) \stackrel {1}{T}(u_1)R_{12}(u_1-u_2),\label{rtt-ttr} 
\eeq
where $\stackrel {1}{X} \equiv  X \otimes 1,~
\stackrel {2}{X} \equiv  1 \otimes X$
for any supermatrix $ X \in End(V) $. For later use, we list some useful
properties enjoyed by the R-matrix:
(i) Unitarity:   $  R_{12}(u)R_{21}(-u) = \rho (u)$ and (ii)
 Crossing-unitarity:  $  R^{st_2}_{12}(-u+1)R^{st_2}_{21}(u) =
         \tilde {\rho }(u)$
with $\rho (u),\tilde \rho (u)$ being some  scalar functions.

In order to describe integrable electronic models on  open
chains, we introduce two associative
superalgebras ${\cal T}_-$  and ${\cal T}_+$ defined by the R-matrix
$R(u_1-u_2)$ and the relations \cite {Zhou97,BRA98} 
\beq
R_{12}(u_1-u_2)\stackrel {1}{\cal T}_-(u_1) R_{21}(u_1+u_2)
  \stackrel {2}{\cal T}_-(u_2)
=  \stackrel {2}{\cal T}_-(u_2) R_{12}(u_1+u_2)
  \stackrel {1}{\cal T}_-(u_1) R_{21}(u_1-u_2),  \label{reflection1}
\eeq
\bea
&&R_{21}^{st_1 ist_2}(-u_1+u_2)\stackrel {1}{\cal T}_+^{st_1}
  (u_1) R_{12}(-u_1-u_2+1)
  \stackrel {2}{\cal T}_+^{ist_2}(u_2)\no\\
&&~~~~~~~~~~~~~~~=\stackrel {2}{\cal T}_+^{ist_2}(u_2) R_{21}(-u_1-u_2+1)
  \stackrel {1}{\cal T}_+^{st_1}(u_1) R_{12}^{st_1 ist_2}(-u_1+u_2),
  \label{reflection2}
\eea
respectively. Here the supertransposition $st_{\alpha}~(\alpha =1,2)$ 
is only carried out in the
$\alpha$-th factor superspace of $V \otimes V$, whereas $ist_{\alpha}$ denotes
the inverse operation of  $st_{\alpha}$. By modifying Sklyanin's 
arguments \cite{Skl88}, one
may show that the quantities $\tau(u)$ given by
$\tau (u) = str ({\cal T}_+(u){\cal T}_-(u))$
constitute a commutative family, i.e.,
        $[\tau (u_1),\tau (u_2)] = 0$. 

One can obtain a class of realizations of the superalgebras ${\cal T}_+$  and
${\cal T}_-$  by choosing  ${\cal T}_{\pm}(u)$ to be the form
\beq
{\cal T}_-(u) = T_-(u) \tilde {\cal T}_-(u) T^{-1}_-(-u),~~~~~~ 
{\cal T}^{st}_+(u) = T^{st}_+(u) \tilde {\cal T}^{st}_+(u) 
  \lt(T^{-1}_+(-u)\rt)^{st}\label{t-,t+} 
\eeq
with
\beq
T_-(u) = R_{0M}(u) \cdots R_{01}(u),~~~~
T_+(u) = R_{0L}(u) \cdots R_{0,M+1}(u),~~~~ 
\tilde {\cal T}_{\pm}(u) = K_{\pm}(u),
\eeq
where $K_{\pm}(u)$, called boundary K-matrices, 
are representations of  ${\cal T}_{\pm}  $ in some representation
superspace. Although many attempts have been made to find c-number
boundary K matrices,which may be referred to as the fundamental
representation,it is no doubt very interesting to search for
non-c-number K matrices,arising as representations in some Hilbert spaces,
which may be interpreted as impurity Hilbert spaces.

We now solve (\ref{reflection1}) and (\ref{reflection2}) 
for $K_+(u)$ and $K_-(u)$. For the quantum R matrix (\ref {r}),
One may 
check that the matrix $K_-(u)$ given by
\beq
K_-(u)=   \left ( \begin {array}
{ccc}
A_-(u)&B_-(u)&0\\
C_-(u)&D_-(u)&0\\
0&0&1
\end {array} \right ),\label{k-}
\eeq
where
\bea
A_-(u)&=&\frac {(c_a-\frac {1}{2})(c_a+\frac {3}{2})-u^2+u+2u {\bf S}^z_a}
{(c_a+u-\frac {1}{2})(c_a+u+\frac {3}{2})},\no\\
B_-(u)&=&\frac {2u {\bf S}^-_a}
{(c_a+u-\frac {1}{2})(c_a+u+\frac {3}{2})},\no\\
C_-(u)&=&\frac {2u {\bf S}^+_a}
{(c_a+u-\frac {1}{2})(c_a+u+\frac {3}{2})},\no\\
D_-(u)&=&\frac {(c_a-\frac {1}{2})(c_a+\frac {3}{2})-u^2+u-2u {\bf S}^z_a}
{(c_a+u-\frac {1}{2})(c_a+u+\frac {3}{2})},
\eea
satisfies (\ref{reflection1}). Here ${\bf S}^{\pm}={\bf S}^x \pm
i{\bf S}^y$.
The matrix $K_+(u)$ can be obtained from the isomorphism of the
superalgebras  ${\cal T}_-  $ and ${\cal T}_+  $. Indeed, given a solution
${\cal T}_- $ of (\ref{reflection1}), then ${\cal T}_+(u)$ defined by
\beq
{\cal T}_+^{st}(u) =  {\cal T}_-(-u+\frac {1}{2})\label{t+t-}
\eeq
is a solution of (\ref{reflection2}). 
The proof follows from some algebraic computations upon
substituting (\ref{t+t-}) into  
(\ref{reflection2}) and making use
of the properties of the R-matrix .
Therefore, one may choose the boundary matrix $K_+(u)$ as 
\beq
K_+(u)=   \left ( \begin {array}
{ccc}
A_+(u)&B_+(u)&0\\
C_+(u)&D_+(u)&0\\
0&0&1 
\end {array} \right ),\label{k-}
\eeq
where
\bea
A_+(u)&=&\frac {c_b^2-u^2-\frac {3}{4} +(2u-1) {\bf S}^z_b}
{(c_b+u+\frac {1}{2})(c_b+u-\frac {3}{2})},\no\\
B_+(u)&=&\frac {(2u-1){\bf S}^-_b}
{(c_b+u+\frac {1}{2})(c_b+u-\frac {3}{2})},\no\\
C_+(u)&=&\frac {(2u-1) {\bf S}^+_b}
{(c_b+u+\frac {1}{2})(c_b+u-\frac {3}{2})},\no\\
D_+(u)&=&\frac {c_b^2-u^2-\frac {3}{4} -(2u-1) {\bf S}^z_b}
{(c_b+u+\frac {1}{2})(c_b+u-\frac {3}{2})}.\no\\
\eea

Now it can be shown  that the 
Hamiltonian (\ref{ham}) is related to the logarithmic derivative of the
transfer matrix
$\tau (u)$ with respect to the spectral parameter $u$ at $u=0$ (up to 
an unimportant additive chemical potential term)
\beq
 -H= \sum _{j=1}^{L-1} H_{j,j+1} + \frac {1}{2} \stackrel {1}{K'}_-(0)
+\frac {str K_+(0)H_{L0}}{str K_+(0)}.
\eeq
This implies that the model under study admits
an infinite number
of conserved currents which are in involution with each other, thus
assuring its integrability.

Having established the quantum integrability of the model,let us now
diagonalize the Hamiltonian (\ref {ham})
by means of the algebraic Bethe ansatz method
\cite {Skl88,Gon94}.
Let us introduce the `doubled' monodromy matrix $U(u)$,
\beq
U(u)=T(u)K_-(u)T^{-1}(-u) \equiv
 \left ( \begin {array}
{ccc}
{\cal A}_{11}(u)&{\cal A}_{12}(u)&{\cal B}_1(u)\\
{\cal A}_{21}(u)&{\cal A}_{22}(u)&{\cal B}_2(u)\\
{\cal C}_1(u)&{\cal C}_2 (u)& {\cal D}(u)
\end {array} \right ).\label{k-}
\eeq
Substituting into the reflection equation (\ref {reflection1}),we may draw the following
commutation relations,
\bea
{\check {\cal A}}_{bd}(u_1){\cal C}_c(u_2)&=&\frac {(u_1-u_2-1)(u_1+u_2)}
{(u_1-u_2)(u_1+u_2+1)}r(u_1+u_2+1)^{eb}_{gh}r(u_1-u_2)^{ih}_{cd}
{\cal C}_e(u_2){\check {\cal A}}_{gi}(u_1)-\no\\
& &\frac {4u_1u_2}{(u_1+u_2+1)(2u_1+1)(2u_2+1)}r(2u_1+1)^{gb}_{cd}
{\cal C}_g(u_1) {\cal D}(u_2) + \no\\
& & \frac {2u_1}{(u_1-u_2)(2u_1+1)}
r(2u_1+1)^{gb}_{id} {\cal C}_g (u_1) {\check {\cal A}}_{ic}(u_2),\label
{cr}\\
{\cal D}(u_1){\cal C}_b(u_2) 
&=&\frac {(u_1-u_2-1)(u_1+u_2)}
{(u_1-u_2)(u_1+u_2+1)}
{\cal C}_b(u_2){\cal D}(u_1)+\frac {2u_2}{(u_1-u_2)(2u_2+1)}
{\cal C}_b(u_1)D(u_2)\no\\
& &  -\frac {1}{u_1+u_2+1}{\cal C}_d(u_1){\check {\cal A}}_{db}(u_2).
\eea
Here $ {\cal A} _{bd}(u) = {\check {\cal A}}_{bd}(u) + \frac {1}{2u+1}
D \delta _{bd}$ and the matrix $r(u)$ ,which in turn satisfies the quantum Yang-Baxter
equation, takes the form,
\beq
r^{bb}_{bb}(u)=1,~~~~~r^{bd}_{bd}=-\frac
{1}{u-1},~~~~~r^{bd}_{db}(u)=\frac {u}{u-1},(b \neq d,b,d = 1,2).
\eeq
 Choosing the Bethe state $|\Omega \rangle $ as
\beq
|\Omega \rangle = {\cal C}_{i_1}(u_1) \cdots {\cal
C}_{i_N}(u_N)|0\rangle F^{i_1\cdots i_N},
\eeq
with $|0\rangle $ being the pseudovacuum, and applying the transfer
matrix $\tau (u)$
to the state $|\Omega\rangle$,we have
$\tau (u) |\Omega \rangle =\Lambda(u) |\Omega \rangle$,with the
eigenvalue,
\bea
\Lambda (u)&=& \frac {2u-1}{2u+1}\frac {(c_b+u-\frac {1}{2})}{(c_b+u+\frac
{1}{2})}
\frac {(c_b+u+\frac {3}{2})}{(c_b+u-\frac
{3}{2})}
(-\frac {u+1}{u-1})^L
\prod ^N_{j=1} \frac {(u+u_j)(u-u_j-1)}{(u-u_j)(u+u_j+1)}\no\\
& &-\frac {2u}{2u+1} (-\frac {u^2}{u^2-1})^L 
\prod ^N_{j=1} \frac {(u+u_j)(u-u_j-1)}{(u-u_j)(u+u_j+1)}
\Lambda ^{(1)}(u;\{u_i\}),
\eea
provided the parameters $\{ u_j\}$ satisfy
\beq
 \frac {2u_j-1}{2u_j}\frac {(c_b+u_j-\frac {1}{2})}{(c_b+u_j+\frac
{1}{2})}
\frac {(c_b+u_j+\frac {3}{2})}{(c_b+u_j-\frac
{3}{2})}
(\frac {u_j+1}{u_j})^{2L}=\Lambda
^{(1)}(u_j;\{u_i\}). \label {bethe1}
\eeq
Here $\Lambda ^{(1)}(u;\{u_i\})$ is the eigenvalue of the transfer
matrix $\tau ^{(1)}(u)$ for the reduced problem,which arises out of the
$r$ matrices from the first term in the right hand side of (\ref {cr}),with
the reduced boundary K matrices $K_{\pm}^{(1)}(u)$ as,
\beq
K^{(1)}_-(u)=
  \left ( \begin {array}
{cc}
A^{(1)}_-(u)&B^{(1)}_-(u)\\
C^{(1)}_-(u)&D^{(1)}_-(u)
\end {array} \right ),\label{k1-}
\eeq
where
\bea
A^{(1)}_-(u)&=&\frac {c_a^{2}-\frac {3}{4}-u^2+(2u+1) {\bf S}^z_a}
{(c_a+u-\frac {1}{2})(c_a+u+\frac {3}{2})},\no\\
B^{(1)}_-(u)&=&\frac {(2u+1) {\bf S}^-_a}
{(c_a+u-\frac {1}{2})(c_a+u+\frac {3}{2})},\no\\
C^{(1)}_-(u)&=&\frac {(2u+1) {\bf S}^+_a}
{(c_a+u-\frac {1}{2})(c_a+u+\frac {3}{2})},\no\\
D^{(1)}_-(u)&=&\frac {c_a^{2}-\frac {3}{4}-u^2-(2u+1) {\bf S}^z_a}
{(c_a+u-\frac {1}{2})(c_a+u+\frac {3}{2})},
\eea
and
\beq
K^{(1)}_+(u)=   \left ( \begin {array}
{cc}
A^{(1)}_+(u)&B^{(1)}_+(u)\\
C^{(1)}_+(u)&D^{(1)}_+(u)
\end {array} \right ),
\eeq
where
\bea
A^{(1)}_+(u)&=&\frac {c_b^2-u^2-\frac {3}{4} +(2u-1) {\bf S}^z_b}
{(c_b+u+\frac {1}{2})(c_b+u-\frac {3}{2})},\no\\
B^{(1)}_+(u)&=&\frac {(2u-1){\bf S}^-_b}
{(c_b+u+\frac {1}{2})(c_b+u-\frac {3}{2})},\no\\
C^{(1)}_+(u)&=&\frac {(2u-1) {\bf S}^+_b}
{(c_b+u+\frac {1}{2})(c_b+u-\frac {3}{2})},\no\\
D^{(1)}_+(u)&=&\frac {c_b^2-u^2-\frac {3}{4} -(2u-1) {\bf S}^z_b}
{(c_b+u+\frac {1}{2})(c_b+u-\frac {3}{2})}.\no\\
\eea
Implementing the change $u \rightarrow u+\frac {1}{2}$ with respect to the
original problem,
one may check that these boundary K matrices satisfy the reflection equations
for the reduced problem. After some algebra,the reduced transfer matrix
$ \tau ^{(1)}(u)$ may be recognized as that for the $(N+2)$-site inhomogeneous
XXX open chain,which has been diagonalized in Ref.\cite {Skl88}. This also
provides an algebraic interpretation for the applicability of the
coordinate Bethe ansatz method \cite {PW97}. Here we merely
give the final result,
\bea
\Lambda ^{(1)}(u;\{ u_j \}) &=&
\frac {(c_b+u-\frac {1}{2})}{(c_b+u+\frac
{1}{2})}
\frac {(c_b+u+\frac {3}{2})}{(c_b+u-\frac
{3}{2})}\prod _{\a =a,b} \frac {u-c_\a-\frac {1}{2}}{u+c_\a+\frac
{3}{2}}
\{ \frac {2u-1}{2u} \prod _{m=1}^M \frac 
{(u-v_m+\frac {3}{2})(u+v_m+\frac {1}{2})}
{(u-v_m+\frac {1}{2})(u+v_m-\frac {1}{2})}\no\\
& &+\frac {2u+1}{2u} \prod _{\a =a,b} \frac {(u-c_\a +\frac
{1}{2})}{(u-c_\a-\frac {1}{2})}
 \frac {(u+c_\a +\frac
{1}{2})}{(u+c_\a-\frac {1}{2})}
\prod _{j=1}^N \frac {(u-u_j)(u+u_j+1)}
{(u-u_j-1)(u+u_j)}
\prod ^{M}_{m=1} \frac {(u-v_m-\frac {1}{2})(u+v_m-\frac {3}{2})}
{(u-v_m+\frac {1}{2})(u+v_m-\frac {1}{2})}\},
\eea
provided the parameters $\{ v_m \}$ satisfy 
\beq
\prod _{\a=a,b} \frac {(v_m+c_\a -1)(v_m-c_\a -1)}
{(v_m+c_\a)(v_m-c_\a)}
\prod ^N_{j=1} \frac {(v_m-u_j-\frac {3}{2})(v_m+u_j-\frac {1}{2})}
{(v_m-u_j-\frac {1}{2})(v_m+u_j+\frac {1}{2})}=\prod ^{M}_{\stackrel {k=1}{k \neq m}}
\frac {(v_m-v_k-1)
(v_m+v_k-2)}
{(v_m-v_k+1)
(v_m+v_k)}. \label {bethe2}
\eeq
After a shift of the parameters $u_j \rightarrow u_j-\frac {1}{2},
v_m \rightarrow v_m + \frac {1}{2}$,the Bethe ansatz equations (\ref
{bethe1}) and 
(\ref {bethe2}) may be rewritten as follows
\bea
(\frac {u_j+\frac {1}{2}}{u_j-\frac {1}{2}})^{2L}
\prod_{\a =a,b}
\frac{u_j+c_\a +1}{u_j-c_\a -1}
& = & \prod_{m=1}^M \frac{u_j-v_m+\frac {1}{2}}{u_j-v_m-\frac {1}{2}}
\frac{u_j+v_m+\frac {1}{2}}{u_j+v_m-\frac {1}{2}},\no\\
\prod_{\a =a,b}\frac{v_m -c_\a-\frac {1}{2}}
{v_m -c_\a+\frac {1}{2}}
\frac{v_m +c_\a-\frac {1}{2}}
{v_m +c_\a+\frac {1}{2}}
\prod_{j=1}^N \frac{(v_m - u_j -\frac {1}{2})}{(v_m-u_j +\frac {1}{2})}
\frac {(v_m + u_j -\frac {1}{2})}{(v_m+u_j +\frac {1}{2})}
   &=&\prod _{\stackrel {k=1}{k \neq m}}
   \frac {(v_m-v_k-1)}{(v_m-v_k+1)}
   \frac {(v_m+v_k-1)}{(v_m+v_k+1)}
  ,\label{Bethe-ansatz}
\eea
with the corresponding energy eigenvalue $E$ of the model 
\beq
E=-\sum ^N_{j=1} \frac {1}{u_j^2-\frac {1}{4}}.
\eeq

In conclusion, we have studied an integrable Kondo problem describing two
impurities coupled to the 1D supersymmetric t-J open chain. 
The  quantum integrability of the
system follows from the fact
that the Hamiltonian may be embbeded into
a one-parameter family of commuting transfer matrices. Moreover, the Bethe
Ansatz equations are derived by means of the algebraic Bethe ansatz
approach. It should be emphasized that the boundary K matices found here
are highly nontrivial,since they can not be factorized into the product
of a c-number K matrix and the local momodromy matrices. However,it is
still possible to introduce a "singular" local monodromy matrix $\tilde
L(u)$
and express the boundary K
matrix $K_-(u)$ as,
\beq
K_-(u)=\tilde {L}(u){\tilde {L}}^{-1}(-u),
\eeq
where
\beq
\tilde L (u) =
 \left ( \begin {array}
{ccc}
u-c_a -1-{\bf S}^z_a&-{\bf S}^-_a&0\\
-{\bf S}^+_a&u-c_a-1+{\bf S}^z_a&0\\
0&0&\e
\end {array} \right ),\label{tl}
\eeq
which constitutes a realization of the Yang-Baxter algebra (\ref
{rtt-ttr}) when $\e$
tends to $0$.
The implication of such a singular factorization deserves further
investigation.
Indeed,this implies that integrable Kondo impurities discussed here
appear to be ,in some sense,related to a singular realization of
the Yang-Baxter algebra,which in turn  reflects a hidden six-vertex XXX
symmetry in the original quantum R matrix. Therefore,one may expect that
the formalism presented here may
be applied to other physically interesting strongly correlated
electron systems,such as the supersymmetric extended Hubbard model and
the supersymmetric U model. Also,the extension of the above
construction to the case of arbitrary impurity spin is straitforward.
The details will be treated in a separate
publication.

\vskip.3in
This work is supported by OPRS and UQPRS. We are grateful to X.-Y. Ge
for enlightening discussions. Thanks are also due to Dr. Y.-Z. Zhang
for his critical comments at the earlier stage of this work.


\end{document}